\documentclass[a4paper]{article}
\usepackage{amsmath}
\usepackage{amssymb}

\newtheorem{theorem}{Theorem}[section]
\newtheorem{lemma}{Lemma}[section]

\newtheorem{remark}{Remark}[section]

\def\Z{{\mathchoice{{\bf Z}}{{\bf Z}}{{\rm Z}}{{\rm Z}}}}

\def\R{{\mathchoice{{\bf R}}{{\bf R}}{{\rm R}}{{\rm R}}}}

\def\T{{\mathchoice{{\bf T}}{{\bf T}}{{\rm T}}{{\rm T}}}}

\let\epsilon=\varepsilon
\let\phi=\varphi
\title{One particle subspaces for two particle quantum walks with ultralocal
interaction}
\author{V.~Malyshev\thanks{Mechanics and Mathematics Faculty, Lomonosov Moscow State University, Leninskie
Gory~1, Moscow, 119991, Russia, E-mail: 2malyshev@mail.ru}
\and
A.~Zamyatin\thanks{Mechanics and Mathematics Faculty, Lomonosov Moscow State University, Leninskie
Gory~1, Moscow, 119991, Russia, E-mail: andrei.zamyatin@mail.ru}
}

\date{}

\begin{document}
\maketitle


\begin{abstract}
We study one particle subspaces for two particles of different masses
with ultra local interaction on a lattice of arbitrary dimension. 
\end{abstract}

\allowdisplaybreaks

\section{Introduction}

We consider here continuous time quantum walks on the integer lattice
$\Z^{d}\subset\R^{d}$. It is well known that for Hamiltonians invariant
w.r.t.\ some translation group there cannot be discrete spectrum.
The convenient and at the same time absolutely rigorous language (among
other approaches) which allows to formulate exactly what means bound
state in this case of $N$ particle problem is the language of 1-,
2-, \ldots  particle subspaces.

Here we consider two particle random walk in any dimension with ultralocal
interaction. Main results are Theorems \ref{t2} and \ref{t3}.

\section{One particle quantum walk}

In this section we will work in the complex Hilbert space $l_{2}(\Z{}^{d})$.
Elements of this space will be denoted as $f=(f(x),x\in\Z^{d})$,
the scalar product is 
\[
(f_{1},f_{2})=\sum_{x\in{\bf Z}^{d}}f_{1}(x)f_{2}^{*}(x).
\]
The standard orthonormal basis consists of the vectors $\delta_{y},$
$y\in\Z^{d}$ such that $\delta_{y}(x)=\delta_{x,y},$ where $\delta_{x,y}=1$
if $x=y$ and zero otherwise.

Define the following linear bounded operators 
\begin{align}
H & =H_{0}+H_{1}:\:l_{2}(\Z^{d})\rightarrow l_{2}(\Z^{d}),\label{eh1}\\
(H_{0}f)(x) & =-\lambda\sum_{k=1}^{d}(f(x+e_{k})-2f(x)+f(x-e_{k})),\label{eh2}\\
(H_{1}f)(x) & =\mu\delta_{x,0}f(x),\label{eh3}
\end{align}
where $e_{k}=(0,\ldots ,0,1,0,\ldots ,0)\in\Z^{d}$ is the unit vector in
the $k$th direction, that is with all zero coordinates except the
$k$th coordinate, $\lambda,\mu$ are real parameters. Moreover, we
assume that $\lambda>0$ as it is standardly accepted, although it
is not really a restriction. It is clear that all these Hamiltonians
are selfadjoint.

Put 
\[
\gamma(\varphi)=\cos\varphi_{1}+\ldots +\cos\varphi_{d}
\]
where $\varphi=(\varphi_{1},\ldots ,\varphi_{d})$ belongs to $d$-dimensional
torus 
\[
\T^{d}=(-\pi,\pi]\times\dots\times(-\pi,\pi].
\]

Below the following integrals are important 
\[
c(d)=\frac{1}{(2\pi)^{d}}\int_{\T^{d}}\frac{d\varphi}{\gamma(\varphi)+d}>0
\]

\[
c_{1}(d)=\frac{1}{(2\pi)^{d}}\int_{\T^{d}}\frac{d\varphi}{\gamma(\varphi)-d}<0
\]
where $d\varphi=d\varphi_{1}\ldots d\varphi_{d}.$

Note that $c(d)=-c_{1}(d).$ In fact, 
\begin{align*}
(2\pi)^{-d}\int_{\T^{d}}\frac{d\varphi}{\gamma(\varphi)+d}&=(2\pi)^{-d}\int_{\T^{d}}\frac{d\varphi}{\gamma(\varphi+\pi)+d}=\\
&=(2\pi)^{-d}\int_{\T^{d}}\frac{d\varphi}{d-\gamma(\varphi)}=-c_{1}(d)
\end{align*}
where $\varphi+\pi=(\varphi_{1}+\pi,\ldots ,\varphi_{d}+\pi).$

These integrals coincide with the classical Watson integral \cite{W}
\[
w(d)=\pi^{-d}\int_{0}^{\pi}\dots\int_{0}^{\pi}\frac{d\varphi}{d-\gamma(\varphi)}.
\]
Namely, 
\[
w(d)=-c_{1}(d)=c(d).
\]
Note that for $d=1,2$ the integral $c(d)=+\infty$, and for $d\geq3$
it is convergent. In the paper \cite{W} integral $c(3)$ was calculated
explicitly. For $d\to\infty$ the asymptotic series is known, see
\cite{JZ}: 
\[
c(d)\thicksim\frac{1}{d}+\frac{1}{2d^{2}}+\frac{3}{4d^{3}}+\dots
\]
Let $\sigma_{ess}(H),$ $\sigma_{p}(H)$ be essential and  point spectra
of $H$ \cite{Simon}.

The following theorem gives complete description of the spectrum.

\begin{theorem}
\label{t1} Let $\lambda>0.$ 
\begin{itemize}
\item For all $\mu$ and all dimensions $d$ we have $\sigma_{ess}(H)=[0,4\lambda d];$ 
\item For $\mu=0$ $\sigma_{p}(H)=\emptyset;$ 
\item For $d=1,2$ and $\mu\neq0$ the  point spectrum $\sigma_{p}(H)$
consists of exactly one eigenvalue $\nu=\nu(\mu,\lambda),$ moreover
$\nu\notin\sigma_{ess}(H);$ 
\item For $d=3,4$ 
\end{itemize}
\begin{quote}
if $|\frac{2\lambda}{\mu}|<c(d)$, then the  point spectrum $\sigma_{p}(H)$
consists of exactly one eigenvalue $\nu=\nu(\mu,\lambda),$ moreover
$\nu\notin\sigma_{ess}(H);$

if $|\frac{2\lambda}{\mu}|\geq c(d)$, then $\sigma_{p}(H)=\emptyset.$ 
\end{quote}
\begin{itemize}
\item For $d\geq5$ 
\end{itemize}
\begin{quote}
if $|\frac{2\lambda}{\mu}|<c(d)$, then the  point spectrum $\sigma_{p}(H)$
consists of exactly one eigenvalue $\nu=\nu(\mu,\lambda),$ moreover
$\nu\notin\sigma_{ess}(H);$

if $\frac{2\lambda}{\mu}=c(d)$, then the  point spectrum $\sigma_{p}(H)$
consists of exactly one eigenvalue $\nu=4\lambda d,$ moreover $\nu\in\sigma_{ess}(H);$

if $\frac{2\lambda}{\mu}=-c(d)$, then the  point spectrum consists
of exactly one eigenvalue $\nu=0,$ moreover $\nu\in\sigma_{ess}(H);$

if $|\frac{2\lambda}{\mu}|>c(d)$, then $\sigma_{p}(H)=\emptyset.$ 
\end{quote}
\begin{itemize}
\item In all cases, if $\mu>0$, then the eigenvalue $\nu\geq4\lambda d;$
if $\mu<0$, then $\nu\leq0;$ equality in these inequalities is achieved
only in the case when $d\geq5$ and $|\frac{2\lambda}{\mu}|=c(d).$ 
\end{itemize}
\end{theorem}

Thus, in dimension $d\geq5$ and if $|\frac{2\lambda}{\mu}|=c(d)$,
one of the boundary points of the essential spectrum belongs to the
 point spectrum. In all other cases the (unique) eigenvalue belongs
to the discrete spectrum (that is, does not belong to the essential
spectrum).

Theorem \ref{t1} is not new -- similar result was proved in \cite{Japan}
for $\mu>0$ and $\lambda=1$, but this not a restriction for their
method. We included this theorem for our paper were self-contained.

\section{Direct integral}

The following definitions are from \cite{Simon} and \cite{Simon4}.

Let $X_{0}$ be a separable Hilbert space, and $(\Omega,\tau)$ measurable
space with $\sigma-$finite measure $\tau$. Vector function $f(\omega):\,\Omega\to X_{0}$
is called measurable, if for any $\xi\in X_{0}$ the function $(\xi,f(\omega))_{X_{0}}$
is measurable. Consider the Hilbert space $X=L_{2}(\Omega,d\tau;\,X_{0})$
of measurable square integrate functions with values in $X_{0}$.
The scalar product 
\[
(f_{1},f_{2})_{X}=\int_{\Omega}(f_{1}(\omega),f_{2}(\omega))_{X_{0}}d\tau,\quad f_{1},f_{2}\in X,
\]
is finite as 
\[
\int_{\Omega}\left\Vert f(\omega)\right\Vert ^{2}d\tau<\infty
\]
Then we will call $X$ the direct integral with the layers isomorphic
to $X_{0}$, and will write 
\[
X=\int_{\Omega}^{\oplus}X_{0}\,d\tau.
\]
Let $\mathcal{L}(X_{0})$ be the space of linear bounded operators
in $X_{0}$. Operator function $A(\omega):\,\Omega\to\mathcal{L}(X_{0})$
is called measurable, if the functions $(\chi,A(\omega)\chi^{\prime})$
are measurable for all $\chi,\chi^{\prime}\in X_{0}.$ Let $L_{\infty}(\Omega,d\tau;\,\mathcal{L}(X_{0}))$
be the space of measurable functions from $\Omega$ to $\mathcal{L}(X_{0})$
such that $ess\,sup\left\Vert A(\omega)\right\Vert <\infty.$

Consider the class of decomposable operators in $X=L_{2}(\Omega,d\tau;\,X_{0})$.
We shall say that linear bounded operator $A:$ $X\to X$ is decomposed
in the direct integral if there exists measurable operator function
$A(\omega)\in L_{\infty}(\Omega,d\tau;\,\mathcal{L}(X_{0}))$ such
that for any $F\in X,$ 
$
(AF)(\omega)=A(\omega)F(\omega).
$
It is commonly written as 
\[
A=\int_{\Omega}^{\oplus}A(\omega)\,d\tau(\omega).
\]
For any $\omega$ the operator $A(\omega)$ will be called the restriction
of the operator $A$ on the corresponding layer.

\section{One particle subspaces for two particle Hamiltonian}

Free one particle quantum walk is defined by the Hamiltonian 
\[
(h_{i}f)(x)=-\lambda_{i}\sum_{k=1}^{d}\left(f(x+e_{k})-2f(x)+f(x-e_{k})\right),\:x\in{\bf Z}^{d},\lambda_{i}\in\R,i=1,2
\]
in the Hilbert space $l_{2}({\bf Z}^{d})$. Then the Hamiltonian for
two non-interacting particles is defined as 
\begin{align*}
H_{0} & =h_{1}\otimes1+1\otimes h_{2}=\\
 & -\lambda_{1}\sum_{k=1}^{d}\left(f(x_{1}+e_{k},x_{2})+f(x_{1}-e_{k},x_{2})-2f(x_{1},x_{2})\right)-\\
 & -\lambda_{2}\sum_{k=1}^{d}\left(f(x_{1},x_{2}+e_{k})+f(x_{1},x_{2}-e_{k})-2f(x_{1},x_{2})\right)
\end{align*}
in the space
$
L=l_{2}({\bf Z}^{d})\otimes l_{2}({\bf Z}^{d})=l_{2}({\bf Z}^{2d})
$
of functions $f(x_{1},x_{2}),(x_{1},x_{2})\in{\bf Z}^{2d}$. We put
for concreteness $\lambda_{1},\lambda_{2}>0$.

We will consider Hamiltonian $H=H_{0}+H_{1}$ for two particles with
the $\delta$-interaction term 
$
(H_{1}f)(x_{1},x_{2})=\mu\delta_{x_{1},x_{2}}f(x_{1},x_{2})
$
where $\mu\in\R$, and $\delta_{x_{1},x_{2}}$ is the Kronecker symbol.
Let $U_{y},$ $y\in{\bf Z}^{d},$ be the translation group in $L,$
$
(U_{y}f)(x_{1},x_{2})=f(x_{1}+y,x_{2}+y).
$
Note that $H$ commutes with $U_{y}$.

The change of variables $x_{1}=x_{1},$ $x=x_{2}-x_{1}$ defines one-to-one
transformation ${\bf Z}^{2d}=\{(x_{1},x_{2})\}\to{\bf Z}^{2d}=(x_{1},x)\}$
and unitary transformation $W_{1}:L=\{f(x_{1},x_{2})\}\to L=\{g(x_{1},x)=f(x_{1},x_{1}+x)\}$.

The translation group now acts only on the first argument: $(U_{y}g)(x_{1},x)=g(x_{1}+y,x)$.

In these coordinates $H$ (in fact $W_{1}HW_{1}^{-1}$) can be written
as follows
\begin{align*}
(Hg)(x_{1},x)= & -\lambda_{1}\sum_{k=1}^{d}\left(g(x_{1}+e_{k},x-e_{k})+g(x_{1}-e_{k},x+e_{k})-2g(x_{1},x)\right)-\\
 & -\lambda_{2}\sum_{k=1}^{d}\left(g(x_{1},x-e_{k})+g(x_{1},x+e_{k}))-2g(x_{1},x)\right)\\
 & +\mu\delta_{x,0}g(x_{1},x)
\end{align*}

Consider the Hilbert space 
\[
\hat{L}=L_{2}(\T^{d})\otimes l_{2}({\bf Z}^{d})
\]
of square integrable functions $F=F(\phi,x),$ where $\phi\in\T^{d},x\in{\bf Z}^{d}.$
The scalar product is defined as 
\[
\langle F_{1},F_{2} \rangle =\sum_{x\in{\bf Z}^{d}}\int_{\T^{d}}F_{1}(\phi,x)\overline{F_{2}(\phi,x)}d\phi.
\]

It is known \cite{Simon}, that the Hilbert space $\hat{L}=L_{2}(\T^{d})\otimes l_{2}({\bf Z}^{d})$
is isomorphic to the space of square integrable functions $l_{2}({\bf Z}^{d})$-valued
functions 
\[
L_{2}(\T^{d},d\phi;\,l_{2}({\bf Z}^{2})),
\]
 where $d\phi$ is the Lebesgue measure on $\T^{d}$.

Thus, according to the above definition, the space $\hat{L}$ can
be represented as the direct integral 
\[
\hat{L}=\int_{\T^{d}}^{\oplus}M\,d\phi
\]
with identical layers $M=l_{2}({\bf Z}^{d}).$

The elements of $\hat{L}$ can also be considered either as complex
functions of two variables $\phi,x$ (then we can denote them as $F(\phi,x)$)
or as functions of one variable $\phi$ with values in the Hilbert
space $l_{2}({\bf Z}^{2})$ (in this case we can use notation $\hat{F}(\phi)$).

Define the linear transformation $\mathcal{F}:\,L\to\hat{L}=L_{2}(\T^{d})\otimes l_{2}({\bf Z}^{d})$
\begin{equation}
(\mathcal{F}g)(\phi,x)=F(\phi,x)=\frac{1}{(2\pi)^{d/2}}\sum_{x_{1}\in{\bf Z}^{d}}g(x_{1},x)e^{i(x_{1},\phi)}\label{f}
\end{equation}
where $\phi=(\phi_{1},\dots,\phi_{d})\in\T^{d}$. Note that $\mathcal{F}$
is a unitary operator. In fact, for any $x\in{\bf Z}^{d}$, by Parseval
equality, 
\[
\sum_{x_{1}\in{\bf Z}^{d}}|g(x_{1},x)|^{2}=\int_{\T^{d}}|F(\phi,x)|^{2}d\phi,
\]
whence 
\[
\sum_{x\in{\bf Z}^{d}}\sum_{x_{1}\in{\bf Z}^{d}}|g(x_{1},x)|^{2}=\sum_{x\in{\bf Z}^{d}}\int_{\T^{d}}|F(\phi,x)|^{2}d\phi.
\]
The adjoint operator $\mathcal{F}^{*}$ is inverse to $\mathcal{F}$
and acts as 
\begin{equation}
(\mathcal{F}^{*}F)(\phi,x)=g(x_{1},x)=\frac{1}{(2\pi)^{d/2}}\int_{\T^{d}}F(\phi,x)e^{-i(x_{1},\phi)}d\phi . \label{in_f}
\end{equation}
Consider the operator $\hat{H}=\mathcal{F}H\mathcal{F}^{*}:\,\hat{L}\to\hat{L}$
which is unitarily equivalent to $H$: 
\begin{multline}
(\hat{H}F)(\phi,x)=\\
=-\sum_{k=1}^{d}\left((\lambda_{1}e^{-i\phi_{k}}+\lambda_{2})F(\phi,x-e_{k})+(\lambda_{1}e^{i\phi_{k}}+\lambda_{2})F(\phi,x+e_{k})\right)+\\
+2d(\lambda_{1}+\lambda_{2})F(\phi,x)+\mu\delta_{x,0}F(\phi,x)\label{h_tilde}
\end{multline}
where $F(\phi,x)\in\hat{L}$.

Define the operator function $\phi\longrightarrow\hat{H}(\phi),$
$\phi\in\T^{d},$ where $\hat{H}(\phi):\,l_{2}({\bf Z}^{d})\longrightarrow l_{2}({\bf Z}^{d})$,
such that 
\begin{align}
(\hat{H}(\phi)u)(x)= & -\sum_{k=1}^{d}\left((\lambda_{1}e^{-i\phi_{k}}+\lambda_{2})u(x-e_{k})+(\lambda_{1}e^{i\phi_{k}}+\lambda_{2})u(x+e_{k})\right)\label{h_hat}\\
 & +2d(\lambda_{1}+\lambda_{2})u(x)+\mu\delta_{x,0}u(x)\nonumber 
\end{align}
for the vector $u=\{u(x),\,x\in{\bf Z}^{d}\}\in l_{2}({\bf Z}^{d}).$
Then $\hat{H}(\phi)$ is measurable in the sense that the scalar product
$(u_{1},\hat{H}(\phi)u_{2})$ is measurable for any $u_{1},u_{2}\in l_{2}({\bf Z}^{d}).$

By (\ref{h_tilde}) and (\ref{h_hat}) the operator $\hat{H}$ can
be represented as 
\begin{equation}
(\hat{H}\hat{F})(\phi)=\hat{H}(\phi)\hat{F}(\phi)\label{m_o}
\end{equation}
where $\hat{F}(\phi)\in\hat{L}.$

It follows from this representation that $\hat{H}$ can be decomposed
in the direct integral 
\[
\hat{H}=\int_{\T^{d}}^{\oplus}\hat{H}(\phi)\,d\phi.
\]
Consider now the spectrum of the operator $\hat{H}(\phi):\,l_{2}({\bf Z}^{d})\to l_{2}({\bf Z}^{d})$
for any fixed $\phi\in\T^{d}$.

Let $\lambda_{1},\lambda_{2}>0.$ Consider the following nonnegative
function of $\alpha\in(-\pi,\pi]$ 
\[
r(\alpha)=\frac{\sqrt{\lambda_{1}^{2}+\lambda_{2}^{2}+2\lambda_{1}\lambda_{2}\cos\alpha}}{\lambda_{1}+\lambda_{2}}>0.
\]
Note that the upper bound in the inequality 
\[
\left(\lambda_{1}-\lambda_{2}\right)^{2}\leq\lambda_{1}^{2}+\lambda_{2}^{2}+2\lambda_{1}\lambda_{2}\cos\alpha\leq\left(\lambda_{1}+\lambda_{2}\right)^{2}
\]
is attained for $\alpha=0$, and the lower bound for $\alpha=\pi$.

Then $|\lambda_{1}-\lambda_{2}|/ (\lambda_{1}+\lambda_{2})\leq r(\alpha)\leq1$
and $r(\alpha)=0\Longleftrightarrow\lambda_{1}=\lambda_{2}\,\&\,\alpha=\pi;$
$r(\alpha)=1$ $\Longleftrightarrow$ $\alpha=0.$ Denote 
\begin{equation}
c(d,\phi)=\frac{1}{(2\pi)^{d}}\int_{\T^{d}}\frac{d\psi}{\sum_{k=1}^{d}r(\phi_{k})(1-\cos\psi_{k})}\label{cdphi}
\end{equation}
where $d\psi=d\psi_{1}\dots d\psi_{d}$ and $\phi=(\phi_{1},\ldots ,\phi_{d})\in\T^{d}.$
If $\phi=\overrightarrow{\pi}=(\pi,\pi,\ldots ,\pi)$ ($\overrightarrow{\pi}$
is $d$-dimensional vector) and $\lambda_{1}=\lambda_{2},$ then the
denominator under the integral equals zero and thus $c(d,\overrightarrow{\pi})=\infty.$

Let $r=\min_{l=1,\ldots ,d}\{r(\phi_{l}):\,r(\phi_{l})\neq0\}>0.$ Denote
$I(\phi)=\{i_{1},\dots,i_{s}\},$ where $1\leq i_{1}<\dots<i_{s}\leq d$
is an array of indices such that $r(\phi_{l})\neq0\Longleftrightarrow l\in I(\phi).$
Here $s=s(\phi)=\#\{r(\phi_{l}),l=1,\ldots ,d:\,r(\phi_{l})\neq0\}.$

Then 
\[
\sum_{k=1}^{d}r(\phi_{k})(1-\cos\psi_{k})=\sum_{k=1}^{s}r(\phi_{i_{k}})(1-\cos\psi_{i_{k}})>r\sum_{k=1}^{s}(1-\cos\psi_{i_{k}})
\]
and 
\begin{align*}
c(d,\phi)&=\frac{1}{(2\pi)^{s}}\int_{\T^{s}}\frac{d\psi_{i_{1}}\dots d\psi_{i_{s}}}{\sum_{k=1}^{s}r(\phi_{i_{k}})(1-\cos\psi_{i_{k}})}\leq\\
&\leq\frac{1}{(2\pi)^{s}r}\int_{\T^{s}}\frac{d\psi_{i_{1}}\dots d\psi_{i_{s}}}{\sum_{k=1}^{s}(1-\cos\psi_{i_{k}})}=c(s)
\end{align*}
Similarly 
\[
c(d,\phi)\geq\frac{1}{(2\pi)^{s}}\int_{\T^{s}}\frac{d\psi_{i_{1}}\dots d\psi_{i_{s}}}{\sum_{k=1}^{s}(1-\cos\psi_{i_{k}})}=c(s)
\]
as $0\leq r(\phi_{l})\leq1.$

Then 
\begin{itemize}
\item for $d=1,2$ $c(d,\phi)$ is divergent for all $\phi\in\T^{d}$; 
\item for $d\geq3$ the integral $c(d,\phi)$ diverges iff $s(\phi)\leq2.$ 
\end{itemize}
The following theorem is a generalization of Theorem \ref{t1}. Put
\begin{align*}
\beta_{1}(\phi) & =2\left(\lambda_{1}+\lambda_{2}\right)\biggl(-\sum_{k=1}^{d}r(\phi_{k})+d\biggr),\\
\beta_{2}(\phi) & =2\left(\lambda_{1}+\lambda_{2}\right)\biggl(\sum_{k=1}^{d}r(\phi_{k})+d\biggr).
\end{align*}
Note that $0\leq\beta_{1}(\phi)\leq\beta_{2}(\phi)\leq4\left(\lambda_{1}+\lambda_{2}\right)d$
and 
\begin{align*}
\beta_{1}(\phi) & =\beta_{2}(\phi)=2\left(\lambda_{1}+\lambda_{2}\right)d\Longleftrightarrow s(\phi)=0\Longleftrightarrow\lambda_{1}=\lambda_{2}\,\&\,\phi=\overrightarrow{\pi}\\
\beta_{1}(\phi) & =0\,\Longleftrightarrow\,\phi=\overrightarrow{0}\\
\beta_{2}(\phi) & =4\left(\lambda_{1}+\lambda_{2}\right)d\,\Longleftrightarrow\,\phi=\overrightarrow{0}
\end{align*}
where $\overrightarrow{0}\in\T^{d}$ is the vector consisting of zeros.

\begin{theorem} \label{t2}
Let $\lambda_{1},\lambda_{2}>0.$ For all $\mu\in\R$, for all dimensions
$d$ and for all $\phi\in\T^{d}$ we have $\sigma_{ess}(\hat{H}(\phi))=\left[\beta_{1}(\phi),\,\beta_{2}(\phi)\right].$

For $\mu=0$ we have $\sigma_{p}(\hat{H}(\phi))=\emptyset$ for all
$\phi\in\T^{d}.$

Let $\mu\neq0.$

For $d=1,2$ and for all $\phi\in\T^{d}$ the  point spectrum $\sigma_{p}(\hat{H}(\phi))$
consists of exactly one eigenvalue $\nu=\nu(\phi,\mu,\lambda_{1},\lambda_{2}),$
where $\nu\notin\sigma_{ess}(\hat{H}(\phi));$

For $d=3,4$ 
\begin{itemize}
\item if condition $\{\lambda_{1}=\lambda_{2}\,\&\,s(\phi)\leq2\}$ holds,
then for all $\phi\in\T^{d}$ the  point spectrum $\sigma_{p}(\hat{H}(\phi))$
consists of exactly one eigenvalue $\nu=\nu(\phi,\mu,\lambda_{1},\lambda_{2}),$
where $\nu\notin\sigma_{ess}(\hat{H}(\phi));$ 
\item if condition $\{\lambda_{1}\neq\lambda_{2}\,\vee\,s(\phi)\geq3\}$
holds, then 
\end{itemize}
\begin{quote}
$\left|\frac{2(\lambda_{1}+\lambda_{2})}{\mu}\right|<c(d,\phi)$ implies
that $\sigma_{p}(\hat{H}(\phi))$ consists of exactly one eigenvalue
$\nu=\nu(\phi,\mu,\lambda_{1},\lambda_{2}),$ where $\nu\notin\sigma_{ess}(\hat{H}(\phi));$

$\left|\frac{2(\lambda_{1}+\lambda_{2})}{\mu}\right|\geq c(d,\phi)$
implies that $\sigma_{p}(\hat{H}(\phi))=\emptyset.$ 
\end{quote}
For $d\geq5$ 
\begin{itemize}
\item if condition $\{\lambda_{1}=\lambda_{2}\,\&\,s(\phi)\leq2\}$ holds,
then for all $\phi\in\T^{d}$ the  point spectrum $\sigma_{p}(\hat{H}(\phi))$
consists of exactly one eigenvalue $\nu=\nu(\phi,\mu,\lambda_{1},\lambda_{2}),$
where $\nu\notin\sigma_{ess}(\hat{H}(\phi));$ 
\item if condition $\{\lambda_{1}=\lambda_{2}\,\&\,s(\phi)=3,4\}$ holds,
then 
\end{itemize}
\begin{quote}
$\left|\frac{2(\lambda_{1}+\lambda_{2})}{\mu}\right|<c(d,\phi)$ implies
that $\sigma_{p}(\hat{H}(\phi))$ consists of exactly one eigenvalue
$\nu=\nu(\phi,\mu,\lambda_{1},\lambda_{2}),$ where $\nu\notin\sigma_{ess}(\hat{H}(\phi));$

$\left|\frac{2(\lambda_{1}+\lambda_{2})}{\mu}\right|\geq c(d,\phi)$
implies that $\sigma_{p}(\hat{H}(\phi))=\emptyset.$ 
\end{quote}
\begin{itemize}
\item if condition $\{\lambda_{1}\neq\lambda_{2}\,\vee\,s(\phi)\geq5\}$
holds, then 
\end{itemize}
\begin{quote}
$\left|\frac{2(\lambda_{1}+\lambda_{2})}{\mu}\right|<c(d,\phi)$ implies
that for all $\phi\in\T^{d}$ the  point spectrum $\sigma_{p}(\hat{H}(\phi))$
consists of exactly one eigenvalue $\nu=\nu(\phi,\mu,\lambda_{1},\lambda_{2}),$
where $\nu\notin\sigma_{ess}(\hat{H}(\phi));$

$\frac{2(\lambda_{1}+\lambda_{2})}{\mu}=c(d,\phi)$ implies that $\sigma_{p}(\hat{H}(\phi))$
consists of exactly one eigenvalue $\nu=\beta_{2}(\phi);$

$-\frac{2(\lambda_{1}+\lambda_{2})}{\mu}=c(d,\phi)$ implies that
the point spectrum consists of exactly one eigenvalue $\nu=\beta_{1}(\phi);$

$\left|\frac{2(\lambda_{1}+\lambda_{2})}{\mu}\right|>c(d,\phi)$ implies
that $\sigma_{p}(\hat{H}(\phi))=\emptyset.$ 
\end{quote}
In all cases the eigenvalue $\nu\geq\beta_{2}(\phi),$ if $\mu>0$
and $\nu\leq\beta_{1}(\phi),$ if $\mu>0.$
\end{theorem}

\begin{remark}
The eigenvalue $\nu=\nu(\phi,\mu,\lambda_{1},\lambda_{2})$ is the
unique solution of the equation 
\begin{eqnarray}
\frac{2(\lambda_{1}+\lambda_{2})}{\mu} & = & \frac{1}{(2\pi)^{d}}\int_{\T^{d}}\frac{d\psi}{\sum_{k=1}^{d}r(\phi_{k})\cos\psi_{k}-d+\frac{\nu}{2(\lambda_{1}+\lambda_{2})}},\label{eq}
\end{eqnarray}
where $d\psi=d\psi_{1}\dots d\psi_{d}.$
\end{remark}

\begin{remark}If $\phi=\vec{0}$ then the Hamiltonian $\hat{H}(\vec{0})$,
defined in (\ref{h_hat}), coincides with the Hamiltonian $H$, defined
in (\ref{eh1})--(\ref{eh3}) if $\lambda=\lambda_{1}+\lambda_{2}$.
Thus, theorem \ref{t1} follows from theorem \ref{t2}.
\end{remark}

Since
\[
0\leq\sum_{k=1}^{d}r(\phi_{k})(1-\cos\psi_{k})\leq\sum_{k=1}^{d}(1-\cos\psi_{k}),
\]
we have
\begin{align*}
c(d,\phi)&=\frac{1}{(2\pi)^{d}}\int_{\T^{d}}\frac{d\psi}{\sum_{k=1}^{d}r(\phi_{k})(1-\cos\psi_{k})}\geq\\
&\geq\frac{1}{(2\pi)^{d}}\int_{\T^{d}}\frac{d\psi}{\sum_{k=1}^{d}(1-\cos\psi_{k})}=c(d)
\end{align*}
and $c(d,\phi)=c(d)$ iff $\phi=\overrightarrow{0}$.

Thus for any $\phi\in\T^{d}$ the operator $\hat{H}(\phi)$ has the
only eigenvalue $\nu=\nu(\phi,\mu,\lambda_{1},\lambda_{2})$ iff one
of the following conditions holds: 
\begin{itemize}
\item $d=1,2$ and $\mu\neq0$ 
\item $d=3,4,$ $\mu\neq0$ and $\left|\frac{2(\lambda_{1}+\lambda_{2})}{\mu}\right|<c(d)$ 
\item $d\geq5,$ $\mu\neq0$ and $\left|\frac{2(\lambda_{1}+\lambda_{2})}{\mu}\right|\leq c(d).$ 
\end{itemize}
From implicit function theorem it follows that $\nu=\nu(\phi,\mu,\lambda_{1},\lambda_{2})$
is a continuous function of $\phi\in\T^{d}.$

Now we give the following fundamental definition. A linear subspace
$L_{1}\subset L$ is called one-particle subspace if 
\begin{itemize}
\item $L_{1}$ is invariant with respect to the translation group $U_{s}$
and with respect to dynamics $e^{itH}$, 
\item there exists vector $g_{0}\in L$ such that $L_{1}$ is generated
by the vectors $\{U_{s}g_{0},s\in{\bf Z}^{d}\}$. 
\end{itemize}
\begin{theorem}\label{t3}Let $\mu\neq0$. Then:

For $d=1,2$ there always exists unique one-particle subspace.

For $d=3,4$ one-particle space exists iff $\,\left|\frac{2(\lambda_{1}+\lambda_{2})}{\mu}\right|<c(d)$.
Then it is unique.

For $d\geq5$ one-particle subspace exists iff $\,\left|\frac{2(\lambda_{1}+\lambda_{2})}{\mu}\right|\leq c(d).$
Then it is unique.
\end{theorem}

Let $x_{1}=(x_{1}^{1},\dots,x_{1}^{d})\in{\bf Z}^{d},$ $x=(x^{1},\dots,x^{d})\in{\bf Z}^{d}.$

\begin{remark}
One-particle subspace is generated by the vectors $\{U_{s}g_{0},s\in{\bf Z}^{d}\}$,
where 
\[
g_{0}(x_{1},x)=\frac{1}{(2\pi)^{d}}\int_{\T^{2d}}\frac{\frac{\mu}{2(\lambda_{1}+\lambda_{2})}\cos x_{1}^{1}\phi_{1}\dots\cos x_{1}^{d}\phi_{d}\cos x^{1}\psi_{1}\dots\cos x^{d}\psi_{d}}{\sum_{k=1}^{d}r(\phi_{k})\cos\psi_{k}-d+\frac{\nu(\phi)}{2(\lambda_{1}+\lambda_{2})}}\,d\phi\,d\psi
\]
belongs to $l_{2}({\bf Z}^{2d})$, and the function $\nu(\phi)$ is
defined as the unique solution of the equation (\ref{eq}).
\end{remark}

\section{Proofs}

\subsection{Proof of Theorem \ref{t1}}

Consider the unitary transformation $U:$ $l_{2}({\bf Z}^{d})\rightarrow L_{2}(\T^{d})$
defined by the one-to-one correspondence between elements of the orthonormal
basis $\delta_{x}$, $x=(x_{1},\dots,x_{d})\in{\bf Z}^{d}$ in $l_{2}({\bf Z}^{d})$
and the elements of the orthonormal basis 
\[
\frac{1}{(2\pi)^{d/2}}e^{i\varphi_{1}x_{1}}\ldots e^{i\varphi_{d}x_{d}},\:(x_{1},\ldots .,x_{d})\in{\bf Z}^{d},
\]
in $L_{2}(\T^{d})$. That is $Uf=F$ where 
\[
f=\sum_{x\in{\bf Z}^{d}}f(x)\delta_{x}\in l_{2}({\bf Z}^{d}),F=\sum_{x\in{\bf Z}^{d}}f(x)\frac{1}{(2\pi)^{\frac{d}{2}}}e^{i\varphi_{1}x_{1}}\ldots e^{i\varphi_{d}x_{d}}\in L_{2}(\T^{d}).
\]
In these terms $\hat{H}=UHU^{-1}:$ $L_{2}(\T^{d})\rightarrow L_{2}(\T^{d})$
can be written as follows 
\[
\hat{H}F=-\lambda(e^{i\varphi_{1}}+e^{-i\varphi_{1}}+\ldots +e^{i\varphi_{d}}+e^{-i\varphi_{d}}-2d)F+\mu f(0)\frac{1}{(2\pi)^{d/2}}.
\]
If for some $\mu$ there exists eigenvalue $\nu$, then the corresponding
eigenfunction $F$ satisfies the equation 
\[
-\lambda(e^{i\varphi_{1}}+e^{-i\varphi_{1}}+\ldots +e^{i\varphi_{d}}+e^{-i\varphi_{d}}-2d)F+\mu f(0)\frac{1}{(2\pi)^{\frac{d}{2}}}=\nu F,
\]
whence 
\begin{align}
F&=\frac{1}{(2\pi)^{d/2}}\frac{\frac{\mu}{2\lambda}f(0)}{\frac{e^{i\varphi_{1}}+e^{-i\varphi_{1}}}{2}+\ldots +\frac{e^{i\varphi_{d}}+e^{-i\varphi_{d}}}{2}-d+\frac{\nu}{2\lambda}}=\nonumber \\[5pt]
&=\frac{1}{(2\pi)^{d/2}}\frac{\frac{\mu}{2\lambda}f(0)}{\gamma(\varphi)-d+\frac{\nu}{2\lambda}}. \label{eq:f(f0)}
\end{align}
For $\mu=0$ from (\ref{eq:f(f0)}) it follows that $F\equiv0$. It
means that for $\mu=0$ there are no eigenvalues.

Note that if $\nu\notin[0,4\lambda d],$ then the denominator in (\ref{eq:f(f0)})
is not zero, and the function $F$ belongs to $L_{2}(\T^{d}).$ As
it is shown in Lemma \ref{l0} (see section \ref{app} below), for
$d\leq4$ $F\notin L_{2}(\T^{d})$ if $\nu\in[0,4\lambda d];$ and
for $d\geq5$ $F\notin L_{2}(\T^{d})$ if $\nu\in(0,4\lambda d)$.

It follows that in dimension $d\leq4$ there are no eigenvalues on
the segment $[0,4\lambda d]$. Similarly, in dimension $d\geq5$ there
are no eigenvalues on the interval $(0,4\lambda d)$.

Let $\mu\neq0$ and $\nu\notin(0,4\lambda d).$ If we expand both
sides of the equality (\ref{eq:f(f0)}) in the basis $(2\pi)^{-d/2}\exp\{i\varphi_{1}n_{1}\}\ldots \exp\{i\varphi_{d}n_{d}\}$,
then all coefficients of both parts should coincide. In particular,
\[
f(0)=\frac{1}{(2\pi)^{d}}\int_{\T^{d}}\frac{\frac{\mu}{2\lambda}f(0)}{\gamma(\varphi)-d+\frac{\nu}{2\lambda}}d\varphi .
\]
Note that as $F$ is not identically zero, then $f(0)\neq0$. We get
then the equation on $\nu$: 
\begin{equation}
\frac{2\lambda}{\mu}=\frac{1}{(2\pi)^{d}}\int_{\T^{d}}\frac{d\varphi}{\gamma(\varphi)-d+\frac{\nu}{2\lambda}}. \label{eq:mnint}
\end{equation}

Consider the case $d=1,2$. Put 
\[
p\left(\nu\right)=\frac{1}{(2\pi)^{d}}\int_{\T^{d}}\frac{d\varphi}{\gamma(\varphi)-d+\frac{\nu}{2\lambda}}
\]
and $\varphi\in T^{d}$. For $\nu>4\lambda d$ the integrand 
\[
\frac{1}{\gamma(\varphi)-d+\frac{\nu}{2\lambda}}>0
\]
is strictly decreasing (if $\nu$ increases) and tends to $0$ as
$\nu\rightarrow+\infty$. It follows that the function $p\left(\nu\right)$
is also strictly decreasing and $p\left(\nu\right)\to0$ as $\nu\rightarrow+\infty$.
As it was mentioned above, for $\nu=4\lambda d$ 
\[
p\left(4\lambda d\right)=\frac{1}{(2\pi)^{d}}\int_{\T^{d}}\frac{d\varphi}{\gamma(\varphi)+d}=+\infty .
\]
Thus, the function $p\left(\nu\right)$ strictly decreases from $+\infty$
to $0$ for $\nu>4\lambda d.$ Then for $\mu>0$ the equation (\ref{eq:mnint})
has a unique solution if $\nu>4\lambda d.$

If $\nu=0$ then $p\left(0\right)=-\infty$. Similarly, we can get
that, if $\nu<0$, $p\left(\nu\right)$ strictly increases from $-\infty$
to zero (as $\nu$ decreases to $-\infty$). That is why for $\mu<0$
the equation (\ref{eq:mnint}) has unique solution for any $\nu<0$.

It follows that the function $p\left(\nu\right)$ takes all values
except zero, moreover exactly once. Then for any $\mu$ there exists
exactly one $\nu$, such that the equation (\ref{eq:mnint}) holds
and there exists unique (up to multiplicative constant) eigenfunction
$F\in L_{2}(\T^{d})$, defined by (\ref{eq:f(f0)}), and such that
$\hat{H}F=\nu F$.

Thus, there exists unique eigenvalue $\nu$ such that $\nu\notin[0,4\lambda d].$
Moreover, $\nu<0$ for $\mu<0$, and $\nu>4\lambda d$ for $\mu>0.$

Let now $d\geq3$. In this case for $\nu=4\lambda d$ and for $\nu=0$
the integrals 
\begin{align*}
p\left(4\lambda d\right)=c(d)= & \frac{1}{(2\pi)^{d}}\int_{\T^{d}}\frac{d\varphi}{\gamma(\varphi)+d}>0\\
p\left(0\right)=-c(d)= & \frac{1}{(2\pi)^{d}}\int_{\T^{d}}\frac{d\varphi}{\gamma(\varphi)-d}<0
\end{align*}
are finite.

Similarly to above, we come to the conclusion that for $\nu\geq4\lambda d$
the function $p\left(\nu\right)$ is strongly decreasing and $p\left(\nu\right)\to0$
as $\nu\rightarrow+\infty,$ and for $\nu\leq0$, if $\nu$ decreases,
the function $p\left(\nu\right)$ strongly increases and $p\left(\nu\right)\to0$
as $\nu\rightarrow-\infty.$ Thus, the function $p\left(\nu\right)$
takes all values except $0$ in the segment $[-c(d),c(d)],$ moreover
each value only once. It follows that for any $\mu$ such that $\left|\frac{2\lambda}{\mu}\right|\leq c(d)$
there is exactly one $\nu$ such that the equation (\ref{eq:mnint})
holds. For $\left|\frac{2\lambda}{\mu}\right|>c(d)$ the equation
(\ref{eq:mnint}) does not have solutions.

If $\mu$ satisfies the condition $\left|\frac{2\lambda}{\mu}\right|<c(d)$,
then the solution $\nu$ of the equation (\ref{eq:mnint}) satisfies
condition $\nu\notin[0,4\lambda d]$, and the function $F$ in the
formula (\ref{eq:f(f0)}) always belongs to $L_{2}(\T^{d})$. Thus,
for such $\mu$ there exists unique eigenvalue $\nu.$

If $\left|\frac{2\lambda}{\mu}\right|=c(d),$ then the solution of
the equation (\ref{eq:mnint}) will be the following: $\nu=0$ for
$\frac{2\lambda}{\mu}=-c(d)$ and $\nu=4\lambda d$ for $\frac{2\lambda}{\mu}=c(d).$

As it follows from lemma \ref{l0}, for $\nu=0,4\lambda d$ the function
$F$ in the formula (\ref{eq:f(f0)}) belongs to $L_{2}(\T^{d})$
only in dimension $d\geq5.$ Whence, for $\mu$ satisfying the condition
$\left|\frac{2\lambda}{\mu}\right|=c(d)$ in dimension $d\geq5$ there
is unique eigenvalue.

\subsection{Proof of Theorem \ref{t2}.}

Fix some $\phi\in\T^{d}$ and let $\nu(\phi)$ be an eigenvalue of
the operator $\hat{H}(\phi)$. Then 
\[
\hat{H}(\phi)\hat{F}_{0}(\phi)=\nu(\phi)\hat{F}_{0}(\phi)
\]
holds where $\hat{F}_{0}(\phi)=\{F_{0}(\phi,x)),\,x\in{\bf Z}^{d}\}\in l_{2}({\bf Z}^{d})$
is an eigenvector corresponding to $\nu(\phi)$.

By (\ref{h_tilde}) and (\ref{m_o}) we get 
\begin{align}
 & -\sum_{k=1}^{d}\left((\lambda_{1}e^{-i\phi_{k}}+\lambda_{2})F_{0}(\phi,x-e_{k})+(\lambda_{1}e^{i\phi_{k}}+\lambda_{2})F_{0}(\phi,x+e_{k})\right)+\label{eig_val_eq}\\
 & +2d(\lambda_{1}+\lambda_{2})F_{0}(\phi,x)+\mu\delta_{x,0}F_{0}(\phi,x)=\nu(\phi)F_{0}(\phi,x),\:x\in{\bf Z}^{d}. \nonumber 
\end{align}

For any $\phi\in T^{d}$ consider the unitary operator $\mathcal{G}:\:l_{2}({\bf Z}^{d})\longrightarrow L_{2}(\T^{d})$
such that 
\[
\mathcal{G}:\:F(\phi,x)\longrightarrow G(\phi,\psi)=\frac{1}{(2\pi)^{d/2}}\sum_{x\in{\bf Z}}F(\phi,x)e^{i(x,\psi)},\;\psi=(\psi_{1},\dots,\psi_{d})\in\T^{d} .
\]
The inverse operator coincides with the adjoint and looks as 
\[
\mathcal{G}^{*}:\:G(\phi,\psi)\longrightarrow F(\phi,x)=\frac{1}{(2\pi)^{d/2}}\int_{\T^{d}}G(\phi,\psi)e^{-i(x,\psi)}d\psi
\]
where $d\psi=d\psi_{1}\dots d\psi_{d}.$ In particular, for $x=0$
\[
F(\phi,0)=\frac{1}{(2\pi)^{d/2}}\int_{\T^{d}}G(\phi,\psi)d\psi .
\]

For any $\phi\in\T^{d}$ the operator $\mathcal{G}\hat{H}\mathcal{G}^{*}$
acts in $L_{2}(\T^{d})$ as follows: 
\begin{align*}
&\bigl(\mathcal{G}\hat{H}\mathcal{G}^{*}G\bigr)(\phi,\psi)=\\
&\quad =-\biggl(\sum_{k=1}^{d}\left((\lambda_{1}e^{-i\phi_{k}}+\lambda_{2})e^{i\psi_{k}}+(\lambda_{1}e^{i\phi_{k}}+\lambda_{2})e^{-i\psi_{k}}
  \! -\! 2(\lambda_{1}+\lambda_{2})\right)\biggr)G(\phi,\psi)+\\
&\qquad {} +\frac{\mu}{(2\pi)^{d}}\int_{\T^{d}}G(\phi,\psi)d\psi=\\
&\quad =-2\biggl(\sum_{k=1}^{d}\left(\lambda_{1}\cos(\psi_{k}-\varphi_{k})+\lambda_{2}\cos\psi_{k}-\lambda_{1}-\lambda_{2}\right)\biggr)G(\phi,\psi)+\\
&\qquad {} +\frac{\mu}{(2\pi)^{d}}\int_{\T^{d}}G(\phi,\psi)d\psi .
\end{align*}
Then the system of equations (\ref{eig_val_eq}) in $L_{2}(\T^{d})$
can be reduced to one equation 
\begin{align*}
&-2\biggl(\sum_{k=1}^{d}\lambda_{1}\cos(\psi_{k}\! -\! \varphi_{k})+\lambda_{2}\cos\psi_{k}\! -\! \lambda_{1}\! -\! \lambda_{2}\biggr)G_{0}(\phi,\psi)
  +\frac{\mu}{(2\pi)^{d}}\int_{\T^{d}}\!\! G_{0}(\phi,\psi)d\psi\\
  &\quad =  \nu(\phi)G_{0}(\phi,\psi)
\end{align*}
where 
\[
G_{0}(\phi,\psi)=\frac{1}{(2\pi)^{d/2}}\sum_{x\in{\bf Z}^{d}}F_{0}(\phi,x)e^{i(x,\psi)} .
\]
Then 
\begin{equation}
G_{0}(\phi,\psi)=\frac{\frac{\mu}{2(2\pi)^{d}}\int_{\T^{d}}G_{0}(\phi,\psi)d\psi}{\sum_{k=1}^{d}\left(\lambda_{1}\cos(\psi_{k}-\varphi_{k})+\lambda_{2}\cos\psi_{k}-\lambda_{1}-\lambda_{2}\right)+\nu/2} .\label{f_tr_1}
\end{equation}
As 
\begin{multline*}
\lambda_{1}\cos(\psi_{k}-\varphi_{k})+\lambda_{2}\cos\psi_{k}=\\
=\lambda_{1}(\cos\phi_{k}\cos\psi_{k}+\sin\phi_{k}\sin\psi_{k})+\lambda_{2}\cos\psi_{k}=\\
=(\lambda_{1}\cos\phi_{k}+\lambda_{2})\cos\psi_{k}+\lambda_{1}\sin\phi_{k}\sin\psi_{k},
\end{multline*}
we have
\[
\lambda_{1}\cos(\psi_{k}-\varphi_{k})+\lambda_{2}\cos\psi_{k}=r^{\prime}(\phi_{k})\cos(\psi_{k}-\eta(\phi_{k}))
\]
where 
\begin{align*}
r^{\prime}(\phi_{k}) & =\sqrt{(\lambda_{1}\cos\phi_{k}+\lambda_{2})^{2}+(\lambda_{1}\sin\phi_{k})^{2}}=\sqrt{\lambda_{1}^{2}+\lambda_{2}^{2}+2\lambda_{1}\lambda_{2}\cos\phi_{k}}\\
\eta(\phi_{k}) & =\arccos\Bigl(\frac{\lambda_{1}\cos\phi_{k}+\lambda_{2}}{p(\phi_{k})}\Bigr) .
\end{align*}
From (\ref{f_tr_1}) we get 
\begin{equation}
G_{0}(\phi,\psi)=\frac{\frac{\mu}{2(\lambda_{1}+\lambda_{2})(2\pi)^{d}}\int_{\T^{d}}G_{0}(\phi,\psi)d\psi}{\sum_{k=1}^{d}r(\phi_{k})\cos(\psi_{k}-\eta(\phi_{k}))-d+\frac{\nu}{2(\lambda_{1}+\lambda_{2})}},\label{f_tr}
\end{equation}
where 
\[
r(\phi_{k})=\frac{r^{\prime}(\phi_{k})}{\lambda_{1}+\lambda_{2}}>0.
\]
For $\mu=0$ from (\ref{f_tr}) it follows that $G_{0}(\phi,\psi)\equiv0$.
It follows that for $\mu=0$ there are no eigenvalues.

Using the evident inequality 
\[
-\sum_{k=1}^{d}r(\phi_{k})\leq\sum_{k=1}^{d}r(\phi_{k})\cos(\psi_{k}-\eta(\phi_{k}))\leq\sum_{k=1}^{d}r(\phi_{k}),
\]
the denominator in (\ref{f_tr}) is never zero for $\nu\notin[\beta_{1}(\phi),\beta_{2}(\phi)]$,
and the function $G_{0}(\phi,\psi)$ evidently belongs to $L_{2}(\T^{d}).$
As it is shown in Lemma \ref{l0} (see section \ref{app} below),
$G_{0}(\phi,\psi)\notin L_{2}(\T^{d})$ for $d\leq4$ and $\nu\in[\beta_{1}(\phi),\beta_{2}(\phi)]$;
and $G_{0}(\phi,\psi)\notin L_{2}(\T^{d})$ when $d\geq5$ and $\nu\in(\beta_{1}(\phi),\beta_{2}(\phi))$.

Thus, in dimension $d\leq4$ there are no eigenvalues $\nu\in[\beta_{1}(\phi),\beta_{2}(\phi)]$,
and in dimension $d\geq5$ there are no eigenvalues such that $\nu\in(\beta_{1}(\phi),\beta_{2}(\phi))$.

Let now $\nu\notin[\beta_{1}(\phi),\beta_{2}(\phi)].$ Integrate both
parts of the equality (\ref{f_tr}) in the vector variable $\psi$:
\begin{multline*}
\int_{\T^{d}}G_{0}(\phi,\psi)\,d\psi=\int_{\T^{d}}G_{0}(\phi,\psi)\,d\psi\times\\
\times\frac{1}{(2\pi)^{d}}\int_{\T^{d}}\frac{\frac{\mu}{2(\lambda_{1}+\lambda_{2})}\,d\psi}{\sum_{k=1}^{d}r(\phi_{k})\cos(\psi_{k}-\eta(\phi_{k}))-d+\frac{\nu}{2(\lambda_{1}+\lambda_{2})}} .
\end{multline*}
Note that $\int_{\T^{d}}G_{0}(\phi,\psi)\,d\psi\neq0,$ otherwise,
by (\ref{f_tr}), we had $G_{0}(\phi,\psi)\equiv0$.

After cancellation we get the following equation for $\nu$: 
\[
1=\frac{1}{(2\pi)^{d}}\int_{\T^{d}}\frac{\frac{\mu}{2(\lambda_{1}+\lambda_{2})}\,d\psi}{\sum_{k=1}^{d}r(\phi_{k})\cos(\psi_{k}-\eta(\phi_{k}))-d+\frac{\nu}{2(\lambda_{1}+\lambda_{2})}} .
\]
From periodicity of the integrand in each variable $\psi_{k}$ it
follows: 
\begin{multline}
(2\pi)^{d}=\int_{-\pi}^{\pi}\dots\int_{-\pi}^{\pi}\frac{\frac{\mu}{2(\lambda_{1}+\lambda_{2})}\,d\psi_{1}\dots\,d\psi_{d}}{\sum_{k=1}^{d}r(\phi_{k})\cos(\psi_{k}-\eta(\phi_{k}))-d+\frac{\nu}{2(\lambda_{1}+\lambda_{2})}}=\\
=\int_{-\pi-\eta(\phi_{1})}^{\pi-\eta(\phi_{1})}\dots\int_{-\pi-\eta(\phi_{d})}^{\pi-\eta(\phi_{d})}\frac{\frac{\mu}{2(\lambda_{1}+\lambda_{2})}d\psi_{1}\dots d\psi_{d}}{\sum_{k=1}^{d}r(\phi_{k})\cos\psi_{k}-d+\frac{\nu}{2(\lambda_{1}+\lambda_{2})}}=\\
=\int_{-\pi}^{\pi}\dots\int_{-\pi}^{\pi}\frac{\frac{\mu}{2(\lambda_{1}+\lambda_{2})}d\psi_{1}\dots d\psi_{n}}{\sum_{k=1}^{d}r(\phi_{k})\cos\psi_{k}-d+\frac{\nu}{2(\lambda_{1}+\lambda_{2})}}\label{similarly}
\end{multline}
That is why $\nu(\phi)$ satisfies the equation 
\begin{equation}
\frac{2(\lambda_{1}+\lambda_{2})}{\mu}=\frac{1}{(2\pi)^{d}}\int_{\T^{d}}\frac{d\psi}{\sum_{k=1}^{d}r(\phi_{k})\cos\psi_{k}-d+\frac{\nu}{2(\lambda_{1}+\lambda_{2})}}\label{eig_val}
\end{equation}
Let us study this equation. Denote 
\[
q(\nu,\phi)=(2\pi)^{-d}\int_{\T^{d}}\frac{d\psi}{\sum_{k=1}^{d}r(\phi_{k})\cos\psi_{k}-d+\frac{\nu}{2(\lambda_{1}+\lambda_{2})}}
\]
For any fixed $\phi$ the function $q(\nu,\phi)$ is defined for $\nu\notin\left(\beta_{1}(\phi);\beta_{2}(\phi)\right)$,
and at the end points of this interval it takes the values 
\[
q(\beta_{2}(\phi),\phi)=c(d,\phi),\:q(\beta_{1}(\phi),\phi)=-c(d,\phi).
\]
In fact, for $\nu=\beta_{2}(\phi)$ we get 
\begin{multline*}
q(\beta_{2}(\phi),\phi)=\int_{\T^{d}}\frac{d\psi}{\sum_{k=1}^{d}r(\phi_{k})(\cos\psi_{k}+1)}=\\
=\int_{\T^{d}}\frac{d\psi}{\sum_{k=1}^{d}r(\phi_{k})(-\cos(\psi_{k}+\pi)+1)}=\\
\int_{\T^{d}}\frac{d\psi}{\sum_{k=1}^{d}r(\phi_{k})(-\cos\psi_{k}+1)}=c(d,\phi)
\end{multline*}
Similarly, for $\nu=\beta_{1}(\phi)$ 
\[
q(\beta_{1}(\phi),\phi)=\int_{\T^{d}}\frac{d\psi}{\sum_{k=1}^{d}r(\phi_{k})(\cos\psi_{k}-1)}=-c(d,\phi).
\]

For fixed $\phi\in\T^{d}$ the function $q(\nu,\phi)$ is strictly
decreasing in $\nu,$ when $\nu\notin\left[\beta_{1}(\phi);\beta_{2}(\phi)\right].$
For $\nu>\beta_{2}(\phi)$ the function $q(\nu,\phi)$ is positive
and tends to $0$ as $\nu\to+\infty.$ For $\nu<\beta_{1}(\phi)$
the function $q(\nu,\phi)$ is negative and tends to $0$ when $\nu\to-\infty.$

Thus, if $c(d,\phi)=+\infty,$ then the function $q(\nu,\phi)$ takes
all real values, except $0,$ and moreover only once, due to strict
monotonicity of the function $q(\nu,\phi)$ in $\nu$ for any fixed
$\phi\in\T^{d}.$ That is why the equation (\ref{eig_val}) has a
unique solution $\nu(\phi).$

If $c(d,\phi)$ is finite, then the function $q(\nu,\phi)$ takes
all values, except $0$, from the finite interval $[-c(d,\phi),c(d,\phi)]$$.$

It follows that there exists exactly one eigenvalue $\nu(\phi)$ for
any $\phi\in\T^{d}$ iff the left hand part of the equation (\ref{eig_val})
belongs to this interval, that is if $|2(\lambda_{1}+\lambda_{2})\mu^{-1}|\leq c(d,\phi)$.

Also it is true that for $\mu>0$ the eigenvalue $\nu(\phi)\geq\beta_{2}(\phi)$,
and for $\mu<0$ we have $\nu(\phi)\leq\beta_{1}(\phi)$.

It remains to check that the eigenfunction, corresponding to $\nu(\phi)$,
belongs to $L_{2}(\T^{d}).$ If the strict inequality $|2(\lambda_{1}+\lambda_{2})\mu^{-1}|<c(d,\phi)$
holds, then $\nu(\phi)\notin\left[\beta_{1}(\phi);\beta_{2}(\phi)\right]$
and the eigenfunction $G_{0},$ defined in (\ref{f_tr}), belongs
to $L_{2}(\T^{d}),$ as the denominator in (\ref{f_tr}) cannot vanish.
If $|2(\lambda_{1}+\lambda_{2})\mu^{-1}|=c(d,\phi)<\infty,$ then
$\nu(\phi)=\beta_{1}(\phi),\beta_{2}(\phi)$ and the eigenfunction
$G_{0}$ belongs to $L_{2}(\T^{d})$ only in dimension $d\geq5.$

The theorem is proved.

\subsection{Proof of Theorem \ref{t3}}

\paragraph{Existence of one particle subspace}

Consider the function $\phi\to\hat{F}_{0}(\phi)=\{F_{0}(\phi,x),\,x\in{\bf Z}^{d}\}\in l_{2}({\bf Z})$,
where $\hat{F}_{0}(\phi)$ is the eigenvector of the operator $\hat{H}(\phi)$,
corresponding to the eigenvalue $\nu(\phi)$. By (\ref{f_tr}), 
\[
  G_{0}(\phi,\psi)=\frac{\frac{\mu}{2(\lambda_{1}+\lambda_{2})(2\pi)^{d/2}}F_{0}(\phi,0)}%
  {\sum_{k=1}^{d}r(\phi_{k})\cos(\psi_{k}-\eta(\phi_{k}))-d+\frac{\nu}{2(\lambda_{1}+\lambda_{2})}}
\]
as 
\[
F_{0}(\phi,0)=\frac{1}{(2\pi)^{d/2}}\int_{\T^{d}}G_{0}(\phi,\psi)d\psi .
\]
Let us find the components $F_{0}(\phi,x),x\in{\bf Z}^{d},$ of the
eigenvector $\hat{F}_{0}(\phi)$, by applying the operator $\mathcal{G}^{-1}$
to the function $G_{0}(\phi,\psi)$: 
\begin{eqnarray}
F_{0}(\phi,x) & = & \frac{1}{(2\pi)^{d}}\int_{\T^{d}}\frac{\frac{\mu}{2(\lambda_{1}+\lambda_{2})}F_{0}(\phi,0)e^{-i(x,\psi)}}{\sum_{k=1}^{d}r(\phi_{k})\cos(\psi_{k}-\eta(\phi_{k}))-d+\frac{\nu}{2(\lambda_{1}+\lambda_{2})}}\,d\psi\qquad \label{gg}\\
 & = & \frac{1}{(2\pi)^{d}}\int_{\T^{d}}\frac{\frac{\mu}{2(\lambda_{1}+\lambda_{2})}F_{0}(\phi,0)e^{-i(x,\psi)}}{\sum_{k=1}^{d}r(\phi_{k})\cos\psi_{k}-d+\frac{\nu}{2(\lambda_{1}+\lambda_{2})}}\,d\psi\label{gf}\\
 & = & \frac{1}{(2\pi)^{d}}\int_{\T^{d}}\frac{\frac{\mu}{2(\lambda_{1}+\lambda_{2})}F_{0}(\phi,0)\cos x^{1}\psi_{1}\dots\cos x^{d}\psi_{d}}{\sum_{k=1}^{d}r(\phi_{k})\cos\psi_{k}-d+\frac{\nu}{2(\lambda_{1}+\lambda_{2})}}\,d\psi\label{g}
\end{eqnarray}
where $x^{k}$ are the coordinates of the vector $x=(x^{1},\ldots ,x^{d})\in{\bf Z}^{d}.$
The equality (\ref{gf}) can be deduced similarly to (\ref{similarly}).
The equality (\ref{g}) holds as 
\[
\int_{\T^{d}}\frac{\sin x^{k}\psi_{k}}{\sum_{k=1}^{d}r(\phi_{k})\cos\psi_{k}-d+\frac{\nu}{2(\lambda_{1}+\lambda_{2})}}\,d\psi=0
\]
because the integrand is odd in the variable $\psi_{k}$.

Denote 
\[
K(\phi,x)=\frac{1}{(2\pi)^{d}}\int_{\T^{d}}\frac{\frac{\mu}{2(\lambda_{1}+\lambda_{2})}\cos x^{1}\psi_{1}\dots\cos x^{d}\psi_{d}}{\sum_{k=1}^{d}r(\phi_{k})\cos\psi_{k}-d+\frac{\nu}{2(\lambda_{1}+\lambda_{2})}}\,d\psi
\]
\begin{equation}
\hat{K}(\phi)=\{K(\phi,x),\,x\in{\bf Z}^{d}\}\in\hat{L}. \label{k}
\end{equation}
Then by (\ref{g}) we have 
\begin{equation}
F_{0}(\phi,x)=F_{0}(\phi,0)K(\phi,x)\Longleftrightarrow\hat{F}_{0}(\phi)=F_{0}(\phi,0)\hat{K}(\phi). \label{g0}
\end{equation}

Introduce the linear subspace $\hat{L}_{1}$ of the space $\hat{L}$
of functions of two variabl\d{e}s $\hat{L}_{1}=\{F(\phi)K(\phi,x),\,F(\phi)\in L_{2}(\T^{d})\}\subset\hat{L}.$
This subspace is isomorphic to $L_{2}(\T^{d})$ and is invariant with
respect to the Hamiltonian $\hat{H}$, where $\hat{H}$ acts in $\hat{L}_{1}$
as the multiplication on the function $\nu(\phi)$: 
\[
\hat{H}:\,F(\phi)K(\phi,x)\longrightarrow\nu(\phi)F(\phi)K(\phi,x).
\]
Put $g_{0}=\mathcal{F}^{*}K\in L$ $L_{1}=\mathcal{F}^{*}\hat{L}_{1}$.
Then $L_{1}$ is one-particle subspace, according to the definition
above, and $L_{1}$ is generated by the vectors $\{g_{s}=U_{s}g_{0},s\in{\bf Z}^{d}\}$.

In fact 
\[
\mathcal{F}^{*}:\,F(\phi)K(\phi,x)\longrightarrow\sum_{s\in{\bf Z}^{d}}f(s)g_{0}(x_{1}-s,x)=\sum_{s\in{\bf Z}^{d}}f(s)U_{s}g_{0}(x_{1},x)\in l_{2}({\bf Z}^{2d})
\]
where 
\[
f(s)=\frac{1}{(2\pi)^{d/2}}\int_{\T^{d}}F(\phi)e^{-i(s,\phi)}\,d\phi .
\]
By (\ref{in_f}) and (\ref{k}) we have 
\begin{align*}
g_{0}(x_{1},x) & =(\mathcal{F}^{*}K)(x_{1},x)=\frac{1}{(2\pi)^{d/2}}\int_{\T^{d}}K(\phi,x)e^{-i(x_{1},\phi)}\,d\phi=\\
 & =\frac{1}{(2\pi)^{3d/2}}\int_{\T^{2d}}\frac{\frac{\mu}{2(\lambda_{1}+\lambda_{2})}e^{-i(x_{1},\phi)}\cos x^{1}\psi_{1}\dots\cos x^{d}\psi_{d}}{\sum_{k=1}^{d}r(\phi_{k})\cos\psi_{k}-d+\frac{\nu}{2(\lambda_{1}+\lambda_{2})}}\,d\phi\,d\psi
\end{align*}
As the integrand is periodic and odd with respect to each variable
$\psi_{k}$ 
\[
\int_{\T^{d}}\frac{\frac{\mu}{2(\lambda_{1}+\lambda_{2})}\sin x^{k}\phi_{k}}{\sum_{k=1}^{d}r(\phi_{k})\cos\psi_{k}-d+\frac{\nu}{2(\lambda_{1}+\lambda_{2})}}\,d\phi=0
\]
Then 
\begin{multline*}
g_{0}(x_{1},x)=\\
=\frac{1}{(2\pi)^{3d/2}}\int_{\T^{2d}}\frac{\frac{\mu}{2(\lambda_{1}+\lambda_{2})}\cos x_{1}^{1}\phi_{1}\dots\cos x_{1}^{d}\phi_{d}\cos x^{1}\psi_{1}\dots\cos x^{d}\psi_{d}}{\sum_{k=1}^{d}r(\phi_{k})\cos\psi_{k}-d+\frac{\nu}{2(\lambda_{1}+\lambda_{2})}}\,d\phi\,d\psi .
\end{multline*}

\paragraph{Unicity of the one particle subspace}

Assume the contrary: that there exists another one-particle subspace
$L_{1}^{\prime}=\{g_{s}^{\prime}=U_{s}g_{0}^{\prime},s\in{\bf Z}^{d}\}$.
Put $\hat{L}_{1}^{\prime}=\mathcal{F}L_{1}^{\prime}.$ This subspace
is generated by the vectors $F_{s}^{\prime}=\mathcal{F}g_{s}^{\prime}=\mathcal{F}U_{s}g_{0}^{\prime}=e^{i(s,\phi)}\mathcal{F}g_{0}^{\prime},$
where $s\in{\bf Z}^{d}.$ It follows that $\hat{L}_{1}^{\prime}$ 
looks like $\hat{L}_{1}^{\prime}=\{F(\phi)K^{\prime}(\phi,x),F(\phi)\in L_{2}(\T^{d})\},$ where
$K^{\prime}=\mathcal{F}g_{0}^{\prime},$ and the function $K^{\prime}$
cannot be presented as 
\[
K^{\prime}(\phi,x)=\nu^{\prime}(\phi)K(\phi,x)
\]
 for some function $\nu^{\prime}\in L_{2}(\T^{d})$, where $K$ is
defined in (\ref{k}). Due to invariance of $\hat{L}_{1}^{\prime}$
with respect to the Hamiltonian $\hat{H}$ we get that for some function
$\nu_{1}(\phi)\in L_{2}(\T^{d})$ the following holds: 
\[
(\hat{H}K^{\prime})(\phi,x)=\nu{}_{1}(\phi)K^{\prime}(\phi,x).
\]
By (\ref{m_o}) this equality is equivalent to 
\[
\hat{H}(\phi)\hat{K}^{\prime}(\phi)=\nu_{1}(\phi)\hat{K}^{\prime}(\phi)
\]
where $\hat{K}^{\prime}(\phi)=\{K^{\prime}(\phi,x),x\in{\bf Z}^{d}\}.$
Remind that $\nu(\phi)$ is the unique eigenvalue of the operator
$\hat{H}(\phi),$ as it was shown above. It follows that $\nu_{1}(\phi)\equiv\nu(\phi)$.
And then by (\ref{k}), (\ref{g0}) for some function $\nu^{\prime}\in L_{2}$
we will have $K^{\prime}(\phi,x)=\nu^{\prime}(\phi)K(\phi,x)$. This
means that subspaces $\hat{L}_{1}$ and $\hat{L}_{1}^{\prime}$ coincide,
what contradicts to our initial assumption.

\appendix
\section{Appendix}
\label{app}

Put 
\[
\gamma_{v}(\varphi)=\sum_{k=1}^{d}v_{k}\cos\varphi_{k}
\]
where $0\leq v_{k}\leq1.$ Consider the integral 
\[
b(y)=\int_{\T^{d}}\biggl(\frac{1}{\sum_{k=1}^{d}v_{k}\cos\varphi_{k}-y}\biggr)^{2}d\varphi
\]
where $y\in[-D,D]$ and $D=\sum_{k=1}^{d}v_{k}.$ If $y\in[-D,D]$,
then the denominator of the integrand can be $0$ and the question
appears whether the integral $b(y)$ is finite or not.

Let $m$ be the number of nonzero coefficients $v_{k}$. Denote $I=\{i_{1},\dots,i_{m}\},$
where $1\leq i_{1}<\dots<i_{m}\leq d$ is the array of indices such
that $v_{l}\neq0\Longleftrightarrow l\in I.$ Without loss of generality
we can assume that $I=\{1,\ldots ,m\},$ where $m\leq d.$ If $m<d$,
then the integrand depends only on the variables $\varphi_{1},\dots,\varphi_{m}$
and 
\begin{multline*}
b(y)=\int_{\T^{d}}\biggl(\frac{1}{\sum_{k=1}^{d}v_{k}\cos\varphi_{k}-y}\biggr)^{2}d\varphi=\\
=(2\pi)^{d-m}\int_{\T^{m}}\biggl(\frac{1}{\sum_{k=1}^{m}v_{k}\cos\varphi_{k}-y}\biggr)^{2}d\varphi_{1}\dots d\varphi_{m} .
\end{multline*}
Thus it is sufficient to consider the integral 
\[
\int_{\T^{m}}\biggl(\frac{1}{\sum_{k=1}^{m}v_{k}\cos\varphi_{k}-y}\biggr)^{2}d\varphi_{1}\dots d\varphi_{m}
\]
where all $v_{k}>0.$

\begin{lemma} \label{l0} 
\begin{enumerate}
\item If $m\leq4$, then $b(\pm D)=+\infty$. If $m\geq5$, then $b(D)<\infty$. 
\item The integral $b(y)$ for $y\in(-D,D)$ is divergent in any dimension. 
\end{enumerate}
\end{lemma}

\textbf{Proof of assertion 1}

Let $y=D$ and 
\begin{equation}
b(D)=\int_{\T^{d}}\left(\frac{1}{\gamma_{v}(\varphi)-D}\right)^{2}d\varphi_{1}\dots d\varphi_{m} . \label{c_d}
\end{equation}
The integrand has singularity only at $\varphi_{1}=\dots=\varphi_{m}=0.$
Consider the integral 
\[
I_{m}=\int_{U_{\delta}}\left(\frac{1}{\gamma_{v}(\varphi)-D}\right)^{2}d\varphi
\]
where $U_{\delta}\subset R^{m}$ is a neighborhood of the point $\varphi_{1}=\ldots =\varphi_{m}=0$
of small radius $\delta$.

From the Taylor expansion $\cos\varphi-1=- \varphi^{2} / 2 +O(\varphi^{4})$
it follows that for sufficiently small neighborhood $U_{\delta}$
of the point $\varphi_{1}=\ldots =\varphi_{m}=0$ we have 
\begin{align*}
I_{m}&=\int_{U_{\delta}}\left(\frac{1}{\gamma_{v}(\varphi)-D}\right)^{2}d\varphi=\\
&=\int_{U_{\delta}}\left(\frac{1}{v_{1}^{2}\varphi_{1}^{2}+\ldots +v_{m}^{2}\varphi_{m}^{2}+O(\varphi_{1}^{4}+\ldots +\varphi_{m}^{4})}\right)^{2}d\varphi .
\end{align*}
First of all we do the change of variables $\varphi_{k}:=v_{k}\varphi_{k}$,
keeping the same notation for the new variable, and then use spherical
coordinates (see, for example, \cite{Shilov}, pp.\thinspace 313) 
\begin{align*}
\varphi_{1}= & r\cos\alpha_{1}\\
\varphi_{2}= & r\sin\alpha_{1}\cos\alpha_{2}\\
\varphi_{3}= & r\sin\alpha_{1}\sin\alpha_{2}\cos\alpha_{3}\\
\dots & \dots\\
\varphi_{m-1}= & r\sin\alpha_{1}\sin\alpha_{2}\dots\sin\alpha_{m-2}\cos\alpha_{m-1}\\
\varphi_{m}= & r\sin\alpha_{1}\sin\alpha_{2}\dots\sin\alpha_{m-2}\sin\alpha_{m-1}
\end{align*}
where $\alpha_{1},\dots,\alpha_{m-2}\in[0,\pi]$ $\alpha_{m-1}\in[0,2\pi].$
The Jacobian of this transformation $J=r^{m-1}\Psi(\alpha_{1},\dots,\alpha_{m-1})=r^{m-1}\sin^{m-2}\alpha_{1}\sin^{m-3}\alpha_{2}\dots\sin\alpha_{m-1}$.

Then 
\[
I_{m}=C_{v}\int_{0}^{\delta}dr\int_{S^{m-1}}\frac{r^{m-1}\Psi(\alpha_{1},\dots,\alpha_{m-1})d\alpha_{1}\ldots d\alpha_{m-1}}{\left(r^{2}+O(r^{4})\right)^{2}}
\]
where $S^{m-1}$ is the $(m-1)$-dimensional sphere of radius $1$
and $C_{v}^{-1}=v_{1}\ldots v_{m}$. Then 
\[
I_{m}=\int_{0}^{\delta}\frac{r^{m-5}}{1+O(r^{2})}dr\int_{S^{m-1}}\Psi(\alpha_{1},\dots,\alpha_{m-1})d\alpha_{1}\ldots d\alpha_{m-1} .
\]
Thus for $m\geq5$ the integral diverges; but for $m\leq4$ it is
finite.

If $y=-D,$ then 
\begin{align*}
b(-D)&=\int_{\T^{m}}\left(\frac{1}{\gamma_{v}(\varphi)+D}\right)^{2}d\varphi=\\
&=\int_{\T^{m}}\left(\frac{1}{v_{1}\cos(\pi+\varphi_{1})+\ldots +v_{k}\cos(\pi+\varphi_{d})+D}\right)^{2}d\varphi=\\
&=\int_{\T^{m}}\left(\frac{1}{-\gamma_{v}(\varphi)+D}\right)^{2}d\varphi=b(D)
\end{align*}

\textbf{Proof of assertion 2}

Let now $y\in(-D,D)$. We shall prove the divergence of the integral
$b(y)$. We shall find point $a\in\T^{m}$ and its neighborhood $V(a)\subset\T^{m}$
so that the integral 
\[
\int_{V(a)}\left(\frac{1}{\gamma_{v}(\varphi)-y}\right)^{2}d\varphi
\]
diverges. All the following is not more than a technical exercise
but it is useful to do it accurately.

For $d=1$ 
\[
b(y)=\int_{-\pi}^{\pi}\left(\frac{1}{v_{1}\cos\varphi-y}\right)^{2}d\varphi .
\]
Take point $a$ such that $\cos a=y/v_{1},$ $-v_{1}<y<v_{1}.$
Then $\sin a\neq0$ and in sufficiently small neighborhood $V(a)$
we have $\cos\varphi-y/v_{1}=\left(-\sin a\right)\left(\varphi-a\right)+O\left((\varphi-a)^{2}\right).$
At the point $a$ the integrand $(v_{1}\cos\varphi-y)^{-2}$
has singularity of the type $(\varphi-a)^{-2}$. That is
why the integral 
\[
\int_{V(a)}\left(\frac{1}{v_{1}\cos\varphi-y}\right)^{2}d\varphi
\]
diverges.

Let $m>1.$ Consider the hypersurface $\Gamma$ 
\[
v_{1}\cos\psi_{1}+\dots+v_{m}\cos\psi_{m}=y,\,-D<y<D .
\]
Choose the point $a=(a_{1},\dots,a_{m})\in\Gamma$ such that all $a_{i}\in(0,\pi)$.
Then $\nabla\gamma_{v}(a)\neq0,$ where $\gamma_{v}(\varphi)=v_{1}\cos\varphi_{1}+\dots+v_{m}\cos\varphi_{m},$
$\varphi=(\varphi_{1},\ldots ,\varphi_{m}).$

Below we shall use the following notation. Vector $\psi=(\psi_{1},\dots,\psi_{m})\in\Gamma$
will denote the corresponding point on the surface, vector $\varphi=(\varphi_{1},\dots,\varphi_{m})\in\T^{m}$
will denote an arbitrary point of the torus $\T^{m}=(-\pi,\pi]\times\dots\times(-\pi,\pi],$
vector $\xi=(\psi_{1},\dots,\psi_{m-1})$ -- coordinates on the surface
$\Gamma.$ Then $\psi_{m}$ is a function of $\xi$ so that $(\xi,\psi_{m}(\xi))\in\Gamma.$

Let $a^{\prime}=(a_{1},\dots,a_{m-1})\in\R^{m-1}.$ Without loss of
generality we can assume that $v_{m}=1.$ In sufficiently small neighborhood
$U(a^{\prime})\subset\T^{m-1}$ of $a^{\prime}$ the surface $\Gamma$
can be defined by the following equations 
\[
\psi_{m}=\psi_{m}(\xi)=\arccos(y-v_{1}\cos\psi_{1}-\dots-v_{m-1}\cos\psi_{m-1})
\]
where $\xi=(\psi_{1},\dots,\psi_{m-1})\in U(a^{\prime}).$

We shall prove that in some neighborhood $V(a)$ of $a$ the integrand
is asymptotically behaves as $c(\varphi) / \rho^{2}(\varphi)$,
where $\rho(\varphi)$ is the distance of point $\varphi\in V(a)$
to $\Gamma$ and $c(\varphi)$ is some smooth function in $V(a)$.
The divergence follows from this.

Let $n(\xi)=\left(n_{1}(\xi),\dots,n_{m}(\xi)\right)$ be the unit
normal to the surface at the point $\psi=(\xi,\psi_{m}),\,\xi\in U(a^{\prime})$
where 
\begin{equation}
n_{i}(\xi)=\left(C(\xi)\right)^{-1}\frac{v_{i}\sin\psi_{i}}{\sqrt{1-\left(y-v_{1}\cos\psi_{1}-\dots-v_{m-1}\cos\psi_{m-1}\right)^{2}}}\label{nor}
\end{equation}
where $i=1,\ldots ,m-1,\:n_{m}(\xi)=\left(C(\xi)\right)^{-1}$ and 
\[
C(\xi)=\sqrt{\frac{\sum_{i=1}^{m-1}v_{i}^{2}\sin^{2}\psi_{i}}{1-\left(y-v_{1}\cos\psi_{1}-\dots-v_{m-1}\cos\psi_{m-1}\right)^{2}}+1}
\]
Let $\psi(\varphi)\in\Gamma$ be the point such that $\varphi$ belongs
to the normal at the point $\psi(\varphi).$

Define $V(a)=\{\varphi:\rho(\varphi)<\delta,\,\psi(\varphi)\in S(a)\},$
where $S(a)=(\psi=(\xi,\psi_{m}(\xi)):\,\xi\in U(a^{\prime}))\subset\Gamma.$
Then $S(a)=V(a)\cap\Gamma.$

In the integral 
\[
\int_{V(a)}\left(\frac{1}{\gamma_{v}(\varphi)-y}\right)^{2}d\varphi
\]
we do the following change of variables 
\[
\varphi_{i}(\xi,r)=\psi_{i}+rn_{i}(\xi),\quad i=1,\ldots ,m-1,
\]
\[
\varphi_{m}(\xi,r)=\arccos(y-v_{1}\cos\psi_{1}-\dots-v_{m-1}\cos\psi_{m-1})+rn_{m}(\xi)
\]
where $\xi=(\psi_{1},\ldots ,\psi_{m-1})\in U(a^{\prime})$ and $r\in I_{\delta}$,
interval of length $2\delta.$ Then 
\[
\varphi(\xi,r)=\psi+rn(\xi)
\]
where $\varphi(\xi,r)=(\varphi_{1}(\xi,r),\dots,\varphi_{m}(\xi,r)),$
$\psi=(\xi,\psi_{m}(\xi))\in\Gamma,$ and 
\[
\psi_{m}=\arccos(y-v_{1}\cos\psi_{1}-\dots-v_{m-1}\cos\psi_{m-1}).
\]

Denote by $J=J(\xi,r)$ the Jacobian of the transformation 
\[
\Phi=(\varphi_{1}(\xi,r),\dots,\varphi_{m}(\xi,r)):\,U(a^{\prime})\times I_{\delta}\to V(a).
\]
We have
\[
\cos\left(\psi_{i}+rn_{i}\right)=\cos\psi_{i}-\left(\sin\psi_{i}\right)rn_{i}+O(r^{2}).
\]
For 
$
y=\sum_{i=1}^{m}v_{i}\cos\psi_{i}
$
and sufficiently small $r$ 
\begin{align*}
y-\sum_{i=1}^{m}v_{i}\cos\varphi_{i}&=\sum_{i=1}^{m}v_{i}\left(\cos\psi_{i}-\cos\left(\psi_{i}+rn_{i}\right)\right)=\\
&=r\sum_{i=1}^{m}v_{i}n_{i}\sin\psi_{i}+O(r^{2}).
\end{align*}
From formula (\ref{nor}) we have 
\[
\sum_{i=1}^{m}v_{i}n_{i}\sin\psi_{i}=\sqrt{\sum_{i=1}^{m}v_{i}^{2}\sin^{2}\psi_{i}}=\left\Vert \nabla\gamma_{v}(\psi)\right\Vert =\left\Vert \nabla\gamma_{v}(\xi,\psi_{m}(\xi))\right\Vert .
\]
Thus the integrand in $V(a)$ can be represented as 
\[
\frac{1}{\left(y-\gamma_{v}(\varphi)\right)^{2}}=\frac{1}{r^{2}\left\Vert \nabla\gamma_{v}(\xi,\psi_{m}(\xi))\right\Vert ^{2}}+O(r^{-4})
\]
and
\begin{multline*}
\int_{V(a)}\left(\frac{1}{\gamma_{v}(\varphi)-y}\right)^{2}d\varphi_{1}\ldots d\varphi_{m}=\\
=\int_{-\delta}^{\delta}r^{-2}\Biggl(\,\int_{U(a^{\prime})}\left\Vert \nabla\gamma_{v}(\xi,\psi_{m}(\xi))\right\Vert ^{-2}J(\xi,r)d\xi\Biggr)dr
\end{multline*}
where $d\xi=d\psi_{1}\dots d\psi_{m-1}.$

As $\left\Vert \nabla\gamma_{v}(a)\right\Vert >0$, we have  in this neighborhood
of $a,$ 
\[
\left\Vert \nabla\gamma_{v}(\xi,\psi_{m}(\xi))\right\Vert >\epsilon>0.
\]
Also 
\begin{multline*}
\int_{-\delta}^{\delta}r^{-2}\Biggl(\, \int_{U(a^{\prime})}\left\Vert \nabla\gamma_{v}(\xi,\psi_{m}(\xi))\right\Vert ^{-2}J(\xi,r)d\xi\Biggr)dr\leq\\
\leq\epsilon^{-2}\int_{-\delta}^{\delta}r^{-2}\Biggl(\,\int_{U(a^{\prime})}J(\xi,r)d\xi\Biggr)dr
\end{multline*}
and we obtain the desired divergence.

\end{document}